\documentstyle[12pt, epsf]{article}
\input epsf
\newcommand{\be}{\begin{equation}}
\newcommand{\ee}{\end{equation}}
\def\bq{\begin{eqnarray}}
\def\eq{\end{eqnarray}}
\def\n{\noindent}

\begin{document}
\pagestyle{empty}

\n
{\bf Seventh Vaidya-Raychaudhari Endowment Award Lecture}

\vspace{5cm}
\n
{\huge {\bf Subtle is the Gravity}}

\vspace{4cm}
\n
{\bf Naresh Dadhich} \\

\n
{\it Inter-University Centre for Astronomy and Astrophysics,}\\
{\it Pune 411 007}.

\vspace{5cm}
\n
{\it Text of the Lecture delivered on 30 January 2001 at the XXI Conference} \\
{\it of Indian Association for General Relativity and Gravitation held at} \\
{\it the Central India Research Institute, Nagpur}.

\newpage
\pagestyle{plain}
\setcounter{page}{1}

\begin{center}
{\huge {\bf Subtle is the Gravity}}
\end{center}

\vspace{1cm}

\begin{center}
{\bf Abstract}
\end{center}

{\bf In this lecture I build up the motivation for relativity and gravitation
based on general principles and common sense considerations which should
fall in the sphere of appreciation of a general reader. There is a novel
way of looking at things and understanding them in a more direct physical
terms which should be of interest to fellow relativists as well as physicists
in general.}    

\vspace{1cm}

 It is indeed a great honour and privelage for me to be asked to deliver the 
Seventh Lecture in the series which not only salutes the seminal and profound 
work of Professors Prahlad Chunilal Vaidya and Amal Kumar Raychaudhuri but 
more importantly also concretises love and affection they enjoy in the Indian 
relativity community. They symbolise  the hardship and perseverance of a 
university academic and are truly the living academic ``dharma''. I thank my 
colleagues on the Selection Committee for giving me this opportunity to pay 
my warmest tributes to them and share with you a viewpoint in understanding 
gravity.

 With Professor Vaidya, I have dual (which is quite characteristic of gravity)
relationship; on one hand we are ``gurubhai'' as students of Professor V. V.
Narlikar and on the other I am his grand-student as a Masters student of 
Professor J. Krishnarao. I thank the latter for setting me on this trail. On 
this occasion I also remember Professor Narlikar with deep sense of 
reverence and gratitude, and I am greatly indebted to him for showing me the 
exciting and amusing way to look at things in one's own way.

 With Professor Raychaudhuri, it is different. There has been no such visible 
agents connecting us but somewhere deep down we seem to share much 
understanding and warmth. To him I have been a naughty student at a 
distance (which helps me to ask stupid questions) and that is how our 
dialogue goes on. After a long trying, I have recently succeeded in coauthoring
a paper with him. That is something I would treasure.  

 Unlike some of my learned and renowned predecessors I do not ask for 
consideration and understanding for inadequacy of what I would say because 
there is no uncertainty and ambiguity about it. It has to be accepted as hard 
fact as such without much ado and pretension. This indulgence on my part 
stems from the belief that if you have  seen a thing differently, then it is 
worth telling about it to others, never mind whether it is right or wrong. 
This is exactly what I intend to do in this lecture and I hope that it would 
not be too taxing for you. \\

\n 
{\bf 1. At the very beginning}

 For the sake of completeness, let us begin at the very beginning with the 
Newtonian framework of classical physics. We have the three laws of motion, 
of which the first makes an equivalence statement that Uniform Motion (motion 
with a constant speed in a straight line) is equivalent to No Motion. That is 
it is impossible to distinguish between these two states by performing any 
physical experiment. Any deviation from this would signal presence or 
application of an external field or force. In a force free region, a particle  
would be in either of these states depending upon the initial condition. In 
here particle is an externally given entity, and no consideration is made 
about how does it come into existence, and what is the energy expended in 
creating it?

 There is an overriding belief bordering to faith that nothing physically 
non-trivial happens without expense of energy and anything that has physical 
existence must have energy. That means the particle that enters in the 
Newtonian physics should also have energy even when it is at rest, let this
be termed as its rest energy. With this bit of extrapolation, energy of a 
particle in free space would be the sum of kinetic energy and its rest 
energy. When particle is at rest, its entire energy is rest energy. How about 
a particle which is never at rest? That is, its entire energy is 
kinetic and rest energy is zero. Its existence is in its motion. Such a 
particle would be moving relative to everyone. With what speed, relative to 
whom, should it move? Since it must move relative to all, hence it must move 
with the same speed relative to all. If there exists such a particle in 
nature, there would exist a universal (invariant) speed which would be 
the limiting speed for all observers. \\

\n
{\bf 2. When there is light}

 Light is that thing which always moves and can never be at rest relative to 
anyone. It is an electromagnetic wave which propagates according to the 
Maxwell theory of electrodynamics. The speed of a wave is entirely determined 
by the properties of the medium in which it propagates. Then its speed can 
only be changed by moving the medium or changing it. If there exists a medium 
which is all pervading, it could neither be moved nor changed. Then the wave 
that propagates in such a medium would have a universal constant speed 
relative to everyone. Such a medium is called vacuum and light propagates in 
it which would hence have the constant invariant speed. A quantum of light is 
that particle, we envisaged earlier, with zero rest energy which cannot be at 
rest relative to any observer.
  
 Thus there does exist in nature something physical that propagates with an
invariant constant speed which is the limiting speed for all material 
particles. The incorporation of this fact would ask for radical modification 
of the Newtonian mechanics. The first obvious casualty would be the law of 
addition of velocities, because anything added to the velocity of light 
should give the same velocity of light. That is $w=u+v$ can no longer remain 
valid, it has to be changed. It should however remain the low velocities 
limit of the new law. Since the light velocity defines an upper threshold 
for propagation of any information, events would now be separated into two 
classes, the ones that are causally connected (whose time separation is 
greater than or equal to the time taken by light to travel from one event 
to the other) and the ones that are causally unconnected (whose time 
separation is less than the light travel time between them). Since there exists
a universal velocity about which all observers agree, it would relate space 
and time and thereby knitting them together into a new entity called 
spacetime. We can measure distance in seconds, time taken by light to travel 
the distance, alternatively we can measure time in centimeters, distance 
traveled by light in the time.

 In the spacetime diagram (it would be insulting for this audience to 
actually draw the diagram), the interior of the light cone indicates 
the causally connected region while the exterior causally unconnected. 
The distance between two locations would measure differently if we travel from 
one to the other by different paths;i.e. distance is a path dependent, as we 
know from everyday experience while traveling by an auto rickshaw/taxi. So 
must be the time between two events because of the equivalence of space and 
time implied by the universal constant speed of light. That is, the time 
between two 
events would dependent upon the path an observer takes to go from one event to 
the other. In principle, the time of the space voyage read off in an 
astronaut's clock would be different from his colleague's on the ground simply 
because they take different paths between the two events, the start and the 
end of 
the voyage. It is a different matter whether this difference is significant 
or not. Thus space and time measurements are relative observer dependent and 
it is the spacetime interval born out of marriage of the two has invariant 
observer independent meaning. ``Space by itself and time by itself are to 
sink fully into shadows and only a kind of union of the two should yet 
preserve autonomy'' thus spoke Minkowski in his famous Cologne lecture in 
1908. \\

\n
{\bf  3. Light binds space and time}

 Newton's first law states that uniform motion is physically undetectable and 
thereby implying equivalence between frames that are in uniform motion for 
all physical phenomena. This is the principle of relativity. Added to this 
 the fact of constancy of velocity of light gives rise to new mechanics which 
 goes by the name, Special Relativity (SR). There is nothing more to it than 
simply deriving the logical consequences of these two basic facts. The salient 
features that emerge are: synthesis of space and time, path dependence of 
time and absence of absolute simultaneity for spatially separated events and 
the quadratic law of conservation of energy and momentum ($E^2 = p^2 + m^2$,   
where $E$, $p$ and $m$ are respectively energy, momentum and rest mass of the  
particle, and the velocity of light $c$ has been set equal to $1$) leading  
to the  equivalence of energy and mass. The relation $E = mc^2$ has for lay  
audience become synonymous to Einstein and is often used to demonstrate the  
immense destructive power encoded in it. Note that conservation law is  
quadratic which distinguishes itself from the Newtonian linear law by the 
fact that a box containing two photons would have in general increase in its  
rest mass due to the presence of photons even though each photon has zero  
rest mass. 

 The frames referred above are the inertial frames (IFs) identified by the  
characteristic that free particles remain either at rest or in uniform  
motion. This is the behavior of free particles in absence of a force or field 
and it defines the standard state of motion. Any deviation from it is  
indicative of presence of a force or field. Force has to be applied  
externally by a mechanical agency while field is a non mechanical agency that  
mediates interaction between particles of specific (charge) property.  

 IFs are connected by the linear Lorentz transformation which keeps the  
velocity of light invariant ensuring its constancy for all IFs. As a matter  
of fact, this transformation is obtained by requiring invariance of the  
velocity of light for any two IFs. That means the universal character of  
velocity of light could be stated in a geometrical form as the invariance of  
spacetime interval defined as follows:
\be 
ds^2 = c^2dt^2 - dx^2 - dy^2 - dz^2 . 
\ee 
It is this spacetime (Minkowski) metric (``distance'') which has the  
invariant observer (IF) independent meaning, and neither spatial distance by  
itself nor time interval by itself. Further it is the description of  
spacetime in absence of all forces, and free particles follow the geodesics  
(straight lines) of this metric. That is motion in absence of forces is  
characterized by geodesics of the Minkowski metric. \\
  
\n
{\bf  4. From flat to curved}  

 One important point that emerges is that any universal physical property or  
statement must be expressible in a geometric form. As does the Minkowski  
metric for velocity of light and free particle motion in absence of force. 
This would serve as a good guiding principle as we go along.  
 
  To measure force we require an IF in actuality. The question is how  
to obtain it in reality? All forces must be removed. There are four basic  
forces in nature, of which the two are short range and hence they do not occur 
 in macroscopic region. The other two are long range forces. Of which the    
electromagnetic force links only to particles having electric charge, and it  
could be shielded off by a suitable prescription for distribution of charges  
and currents. On the other hand the remaining force of gravitation has  
universal linkage and it therefore cannot be shielded off. Newton too did  
recognise the difficulty in removing gravity, and had prescribed to go  
infinitely away from matter distribution to actually realise an IF.  

 Since gravity cannot be removed globally, there cannot exist a global IF.  
However it can be removed locally as demonstrated by Galileo's famous 
experiment of the leaning tower of Pisa. All particles share the two 
properties, one inertia and other linkage to gravity (gravitational charge). 
The only one universal parameter available is mass (energy) for all 
particles. Thus the measure of inertia and gravitational charge must be the 
same, which would mean all particles irrespective of their mass, shape and 
substance would fall with the same acceleration under gravity. This was what 
for the first time experimentally tested by Galileo. In Einstein's words, 
gravity could be removed by letting oneself fall along with a freely falling 
lift. In contrast to Galileo, Einstein fortunately did this experiment only 
in thought and not in actuality, else he won't have been there to tell us the 
result and the following story of general relativity.  

 Freely falling lifts would thus define true IFs but only locally. That is,  
we can construct a LIF anywhere in spacetime but no GIF. The principle of  
relativity would now state that all LIFs are equivalent, and it is given the  
sophisticated name, Principle of Equivalence (PE). That is, by doing any  
physical experiment in actuality or in thought it is impossible to distinguish 
 one LIF from the other. Conversely, gravity could be locally produced as well 
 by giving upward acceleration equal to that of gravity to a lift in free  
space. This situation is indistinguishable from that of the inside of a lift  
at rest, say on the surface of earth. The strong form of PE states that all  
LIFs irrespective of their location in space and time would measure the same  
constant numerical values of the universal constants, the velocity of light  
$c$, the gravitational constant $G$ and the Planck constant $h$. 

 The non existence of GIF has profound consequence that globally spacetime 
cannot be described by the Minkowski metric, which is flat having zero Riemann 
curvature. This clearly means that for honest and proper consideration of 
gravitational field, we will have to give up the comfort and ease of the non 
interacting inert flat spacetime typified by the Minkowski metric. This is a 
radical break. No other force interacts with background spacetime while 
gravity, it appears, cannot do without it. In the local neighbourhood we can 
always define a LIF, and write the flat Minkowski metric while globally the 
metric is non flat. That means spacetime must be curved in such a way that 
the effect of curvature is locally ignorable permitting existence of a LIF 
with the Minkowski metric. Clearly it must be a Riemannian manifold with 
the characteristic property of existence of a tangent plane defining a LIF.
Note that the characterizing property for flat space is vanishing of the 
curvature rather than the form of the metric (the metric coefficients could 
be functions of coordinates yet the metric could be flat). This is to  
distinguish between globally removable inertial forces (centrifugal and 
coriolis) by coordinate transformation, and globally non removable force of 
gravitation.  

 It is the non removability of gravity which is responsible for curvature of 
the spacetime with non-Minkowskian metric given by
\be
ds^2 = g_{ab}(x^i)dx^adx^b . 
\ee
Free particle motion would be given by the geodesic of this metric which 
would correspond to motion under gravity. Thus gravity is completely 
incorporated in the spacetime geometry and is described by curvature of the 
spacetime manifold. This leads to Einstein's theory of gravitation, called 
general relativity (GR). We have now to obtain the gravitational field 
equation from the curvature of spacetime. \\
  
\n
{\bf  5. Light bends space}

 In addition to the above strong motivation for curved spacetime for 
description of gravitation, here is yet another very strong and inescapable 
motivation. With the equivalence of mass and energy, energy in any form must 
do all that mass does. That means all that which has energy in any form must 
interact with gravitational field. How is particle's interaction with any 
field determined? By looking at the change in its velocity. Light however 
cannot by definition suffer any change in its velocity. Then how could it 
feel gravity? And that it must. In the Newtonian theory, a free particle 
instantaneously at rest experiences gradient of potential as acceleration 
which causes change in its velocity. Since photon (light) is never at rest 
relative to any frame, it cannot experience the gravitational force through 
gradient of potential. 

 This clearly brings forth the fact that if light has to interact with gravity,
 we will have to do something radically different. The pertinent consideration
 would then be what is it that light can respond to ?

 Light propagates as a wave in space, hence it could only respond to changes 
in space.  By 
light's interaction with gravity, what is expected is that it should bend like 
other particles rather than traveling along a straight line. This can 
alternatively be achieved by bending the space rather than light. 
The only sensible thing would then be that gravity must curve space
 and light simply propagates along the geodesic of the curved space.
The physical
 consistency of all the principles involved thus demands that gravitational 
field must curve space. That is, a gravitating massive body curves space 
around it. With SR behind us, curvature of space should also imply curvature 
of spacetime. In fact, it is $g_{tt}$ in the metric which generates 
$-\nabla \phi$ in the geodesic equation for ordinary particles. While for 
photon both $g_{rr}$ and $g_{tt}$ contribute equally in the 
``space bending'' (note that the Newtonian bending value was half of the 
correct value because it referred only to $g_{tt}$). It is the space that 
bends and not the light. Note however that the Weyl curvature of 4-spacetime 
must be non zero to make light feel the bending of space. What was for the 
first time measured by Eddington's team during the total solar eclipse of 
1919 was not the bending of light but was in fact the bending of space. It 
is something like we see the sun going from east to west, but in fact it is 
the earth that is moving about the sun.

  Gravitation could thus be described only through curvature of spacetime, and 
the theory that does this is GR. \\

\n
{\bf  6. Another view}

 Let us once again begin at the very basics. For existence of any physical 
entity which we would call by the general name particle, the minimal 
requirement is that it possesses some energy in some form. This is the bare 
situation. Such an entity could be characterized by a single parameter, the 
measure of its energy, which could be in the form of rest mass or purely 
kinetic like that of a photon. This characteristic would however be common 
for all other particles possessing additional characteristics like electric 
charge. This is thus a universal property of all particles.

 All particles must interact with each other according to their 
characteristics. That means there must exist a universal interaction which 
is shared by all things. Since this is universal and would hence influence 
one and all irrespective of their location in space and time. The interaction 
must therefore be long range. By the term universal we would mean interaction 
with all forms of energy as well as at everywhere in spacetime. Such a 
universal 
field could not be removed globally so long as there exist some particles, in 
the limit at least one (at least two if you adhere to Mach's Principle), in 
the Universe. That is, it can never be completely and absolutely empty.

 As argued earlier, because of the non removability of the universal 
interaction, there cannot exist a GIF and thus spacetime cannot be described 
by the Minkowski metric. Since the Minkowski metric provides a good 
background for the rest of physical phenomena, it must be available locally 
in a laboratory. That means there must exist LIFs in a local neighbourhood. 
In view of this, the metric should thus be written in such a way that it 
represents a curved Riemannian 4-spacetime manifold. The cause for curvature 
is the universal interaction which is shared by all particles. The 
universality demands that it must be synthesized in the spacetime geometry. 
It should then become simply a property of spacetime and is completely 
described by its curvature. 

 In any spacetime including the Newtonian space, the equation of motion of 
free particle is always free of mass of the particle. This is just to indicate 
that free particle motion is always determined by the geometry of 
spacetime/space. Free particle motion would thus be given by the geodesics of 
the curved spacetime metric. The geodesic equation is free of particle's mass.
 Since the interaction is universal, it cannot distinguish one particle from 
the other on the basis of mass. Another reason for absence of mass in the 
equation is that we also have particles with zero (rest) mass, and they are 
always free. This property is characteristic of the equation being determined
by the spacetime/space as its geodesic.

 We have thus a curved Riemannian spacetime of 4-dimensions which imbibes in 
its curvature the universal interaction. How do we deduce 
the equation of motion for the interaction? It is the curvature of spacetime 
which should yield the equation. The Riemann 
curvature satisfies the Bianchi differential identity, which on contraction 
yields the divergence free second rank symmetric tensor, called the Einstein 
tensor $G_{ab}$. The natural equation that could thus follow is:
\be
G_{ab} = -\kappa T_{ab} + \Lambda g_{ab}
\ee
where $\kappa$ and $\Lambda$ are constants and $T_{ab}$ is a symmetric tensor 
with vanishing divergence. This requirement ensures conservation of what the 
symmetric tensor $T_{ab}$ may represent. The natural choice for that is the 
energy-momentum tensor of a matter distribution. The second term on the right 
is a constant for the covariant derivative relative to the curved metric 
$g_{ab}$. 
                                                                              
 The question is, could we make any sense out of this equation? At the outset,
it looks a good equation ensuring the conservation of energy and momentum 
represented by the tensor $T_{ab}$, and $\Lambda$ could be identified with the 
well-known cosmological constant. Let us solve it for $T_{ab} = \Lambda = 0$ 
for a radially symmetric metric. It is most remarkable that the solution 
includes in the first approximation  gravitational field of a mass point in 
the Newtonian theory. This implies that the universal interaction is that of 
gravitation and the above equation is indeed Einstein's equation for 
gravitation in GR.  
 
 We have thus obtained Einstein's equation from a very general consideration 
which entirely hinges on universality. Isn't it amazing that it leads to a 
solution which agrees with the specific inverse square law of gravitation? 
Recall that the demand of existence of closed orbits for the central force 
does pick out the inverse square law of gravitation or electrostatics and the 
linear law of simple harmonic oscillator. Further demand of long range 
isolates the inverse square law. 

 How did this specific property get incorporated in our general consideration?
Our main plank was universality which lead to curved spacetime and the above 
equation. The next step was to identify the energy momentum tensor of 
matter distribution with the tensor $T_{ab}$. This identification ensured 
conservation of energy and momentum. Of course it is well known that the 
Gauss law which implies the inverse square law is essentially a conservation 
law. It is perhaps the implied 
conservation law in our consideration that makes the bridge with the 
Newtonian gravity. At any rate, it is quite remarkable that the 
universality of interaction alone leads uniquely to Einstein's theory of 
gravitation. 

 From this standpoint, the view that spacetime should incorporate gravitation 
in its geometric structure is quite natural because of its universal character
 and it needs no further justification. Newtonian gravity is not quite 
universal (it was though universal enough for its times) as it does not 
include photons. We wish to say that GR results naturally and uniquely when 
the Newtonian theory is rendered truly universal. That is the complete 
universalization of the Newtonian gravitation is Einstein's theory of 
gravitation. This is the fundamental statement. \\ 
 
\n
{\bf 7. A quadratic force} 

 The Lorentz force on a charged particle in electromagnetic field is a linear 
force law;i.e. it is linear in 4-velocity of the particle. Recently we have 
attempted to obtain the entire set of the Maxwell equations by demanding the 
force to be linear in 4-velocity in the relativistic law of motion. Apart 
from this we required the equation to follow from a Lagrangian, and there is 
no a priori choice between scalar and pseudo scalar charges. Then the 
solvability of the system ultimately leads to the Maxwell equations [1]. The 
question naturally arises, what would the quadratic law lead to? Perhaps to 
the Einstein equation, because gravitation is in a sense generalization of the 
electromagnetic field. That can only happen if the quadratic force requires 
the spacetime metric as its potential.

 The relativistic law of motion is, $mdu^a/ds = f^a$, and we wish to consider 
the quadratic force, $f^a = T^a_{bc}u^bu^c$. The Lagrangian for a quadratic 
force would also have the quadratic term of the kind, $p_{ab}(x^i)u^au^b$. 
This would on variation give the equation,
\be
p_{ab}du^b/ds = - P_{(bc,a)}u^bu^c
\ee
where
\be
P_{(bc,a)} = \frac{1}{2} (p_{ba,c} + p_{ac,b} - p_{bc,a}).
\ee 
A comma denotes partial derivative. 

 The first point to note is that the equation is free of mass of the particle.
This is the characteristic property of the quadratic force. This is an 
important inference which implies that the quadratic force would 
automatically obey PE. The equation when it is free of mass of 
the particle, it should be given by the geodesic of the spacetime metric.

 The 4-force should be orthogonal to 4-velocity requiring $p_{ba,c}u^bu^cu^a 
= 0$ which is impossible as $p_{ab}$ is symmetric. The equation of motion 
could make sense if and only if $p_{ab} = g_{ab}$ defines the spacetime 
metric. 
Then it would become the geodesic equation, indicating vanishing of the 
4-acceleration relative to the metric $g_{ab}$ and which would by definition 
be orthogonal to 4-velocity. The quadratic force thus requires spacetime 
metric as its potential which means it gets incorporated in spacetime 
geometry. This 
is the distinguishing and unique feature. The quadratic force comes 
through the Christoffel symbols of the metric, which are not tensors and 
involve first derivatives of the metric. The force could thus always be 
removed globally if space is flat but only locally if space is curved. There 
are two quadratic forces known to us; the inertial (centrifugal and coriolis)
forces belong to the former category of global removability while gravitation 
belongs to the latter category of only local removability and of global non 
removability.  

 For the quadratic force to be universal that it is globally non removable, 
spacetime must be a curved Riemannian manifold. Once spacetime is curved, 
the Einstein equation 
could be obtained as done above in 3. The remarkable point is that the very 
general consideration of force being quadratic leads uniquely to the Einstein 
equation. 
Note that PE follows as a consequence of the quadratic law and not assumed a 
priori. The quadratic law thus characterizes mass proportional force fields 
and makes the spacetime metric dynamic when the force is globally non 
removable. \\

\n
{\bf 8. Some Subtelities}

 Newtonian theory was remarkably successful in explaining motion of planets 
in the solar system, and occurrence of eclipses and other such observable 
phenomena were predicted and verified at astonishingly high degree of 
precision. This raises a simple question what was then wrong with it which did 
not show up in these observations? The first and foremost in this context was 
interaction of light with gravity. This was not only difficult to observe but 
 one did not even attempt to look for it as the theory did not predict it. It 
only became pertinent after the advent of SR. \\

\n
 8.1 {\it Gravitational Potential}

 New theory always includes the old theory and one understands the former from
the framework of the latter. It is therefore pertinent to consider the 
question, how does the new theory manifest in relation to the Newtonian 
theory? The Newtonian gravity is described by a scalar field satisfying the 
Laplace/Poisson equation, $\nabla^2\phi = 0,\rho$; where $\rho$ denotes the 
matter density. The motion under gravity was described by the equation, 
$\bf{\ddot x = - \nabla}\phi$, which is free of mass of the particle by 
the prescription that gravitational force is mass proportional. The two main 
objections are: One, the former equation does not describe how the field 
propagates in space? Second, since gravitational field also has energy, which 
should also participate in gravitational interaction. That is the self 
interaction, gravity being its own source. This is something different and 
unique to gravity. Even in the empty space which is free of ponderable 
matter/energy, gravitational field energy would be present. Thus the correct 
equation for empty space should be $\nabla^2\phi = - \frac{1}{2}(\nabla\phi)^2$, a 
non linear differential equation. 

 To incorporate the particle equation of motion in the geodesic 
equation, it is sufficient to write $g_{tt} = 1 +2\phi$. This is how the 
Newtonian potential enters into the metric. The gravitational field equation 
is as given by the equation (3). It is a second order non 
linear differential equations for the ten metric potentials which have four 
degrees of coordinate freedom implied by the Bianchi identities. The equation 
for empty space is obtained by setting $T_{ab} = 0 = \Lambda$, which is the 
analogue of the Laplace equation. It reads as $G_{ab} = R_{ab} - \frac{1}{2} R \, g_{ab} = 
0$ which is equivalent to the Ricci tensor, $R_{ab} = 0$.

 First of all let us enquire, do we solve the Laplace equation or the non 
linear equation containing the field energy density? Let us solve the vacuum 
equation for the field of a mass point. Consider the spherically symmetric 
metric given by
\be
ds^2 = A dt^2 - B dr^2 - r^2(d \theta^2 + sin^2\theta d\varphi^2)
\ee
where $A$ and $B$ are functions of $r$ and $t$. We are to solve the equation 
$R_{ab} = 0$.

 It turns out that $R_{rt} = 0$ and $R^t_t = R^r_r$ would imply $A = 1/B = 
1 + 2\phi(r)$, and then,
\be
R^t_t = -\nabla^2\phi
\ee
and
\be
R^\theta_\theta = R^\varphi_\varphi = -\frac{2}{r^2}(r\phi)^\prime
\ee
where prime denotes derivative relative to $r$.

 Note that we again solve the good old Laplace equation in $R^t_t = 0$ and not 
the one with the field energy density. Secondly, the other equation is its 
first integral which fixes the constant of integration to zero, determining 
the potential $\phi = -M/r$ absolutely, which can vanish only at infinity and nowhere else. And we obtain the asymptotically flat Schwarzschild solution. Note 
that asymptotic flatness was not due to the boundary condition but instead 
implied by the equations themselves.

 Thus gravitational potential in GR is determined absolutely as 
constant potential would produce non-zero curvature, $R^\theta_\theta$, 
and hence would have physical meaning. This is something new in contrast to 
the classical physics where constant potential is physically ignorable 
because it had no physical effect. 

 Since we ultimately solve the Laplace equation, what has happened to the 
contribution of field energy? Though field energy must participate but it is 
a second order source and is not equivalent to the primary source of 
non-gravitational energy distribution. How could we distinguish between them? 
In the 
classical framework there exists no sensible way of doing it. 
Once again we are at the threshold of 
some new and novel construct.
The only 
natural way is to curve the space so that contribution of the field energy is 
taken care of by its curvature leaving the Laplace equation intact. This is 
exactly what happens.  If we consider space to be flat by setting $B = 1$, 
the equation $R^t_t = 0$ would have the non linear term of the field energy.  
It is the condition $A = 1/B$ which kicks out the non linear term and leaves 
only the Laplacian. Thus contribution of field energy goes into curving the 
space. 

 Since the gravitational potential $\phi = -M/r$ has very strong observational 
support, it couldn't have been tampered with. Here again space is required to 
be curved to account for the field energy's contribution as was the case for 
negotiating light's interaction with gravitation. Both these are non-Newtonian
 effects. It is interesting to note that the non-Newtonian effects would 
primarily manifest through the space curvature. \\

\n
 8.2 {\it Negative Curvature}

 There are theorems establishing positivity of non-gravitational energy, 
which produces the attractive gravitational field, whose energy would always 
be negative. The positive energy condition could alternatively be stated 
as the field energy being negative. This feature must be reflected in the 
space curvature which is caused by the field energy. The potential gradient 
pulls free particles towards the gravitating source, space curvature must  
also act in line with this motion. 

\begin{figure}
\centering
{\epsfysize=6.5cm
{\epsfbox[0 0 500 250]{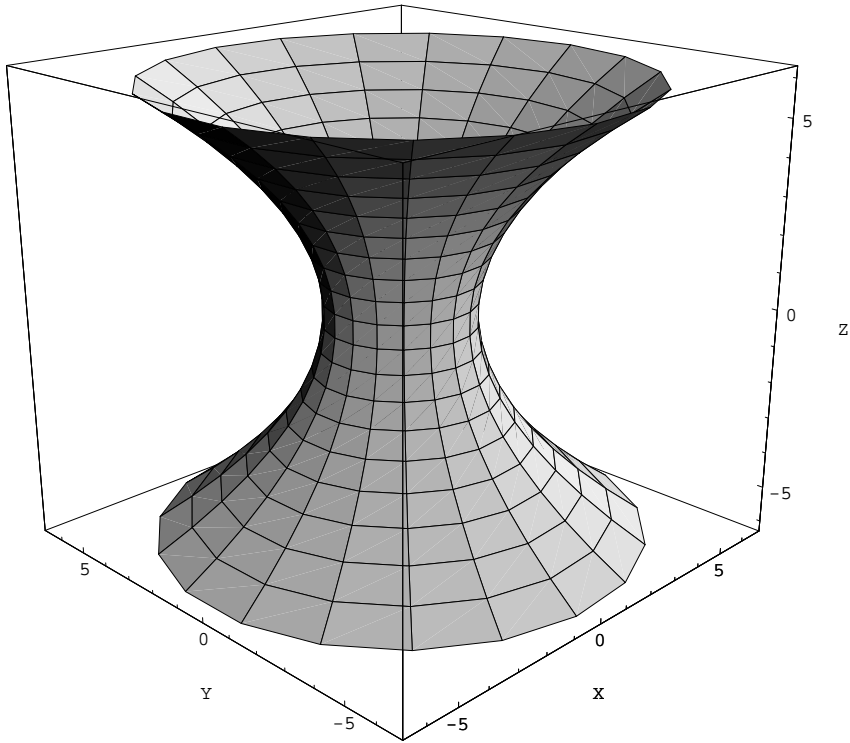}}} 
\caption[]{It shows the paraboloid of revolution for the parabola
 $z^2 = 8 r - 16$, where $r^2 = x^2 + y^2$.}
\end{figure}

 Let us try to understand this phenomenon with the help of the Schwarzschild 
solution. For motion in the equatorial plane, the 2-space metric is given by
\be
(1 - 2M/r)^{-1}dr^2 + r^2d\varphi^2
\ee
 which has the negative curvature $-M/r^3$. It can be embedded into the 
3-Euclidean space by writing $z^2 = 8M(r - 2M)$, which is a parabola and 
would generate a paraboloid of revolution (Fig.1). Clearly it has negative 
curvature which would tend free particles to roll down towards the centre and 
thus work in unison with the potential gradient [2].

Gravitational field energy thus produces negative curvature, and this would 
also be the norm for non-gravitational energy which is non-localizable. That 
is the proper energy condition for non-localizable energy distribution is 
that it be negative so as to work in unison with the localized 
energy. For instance, non-localized energy distribution for the charged black 
hole solution (the electric field energy) and also for the Schwarzschild - de 
Sitter solution is positive and hence it produces repulsive contribution in 
both cases opposing the attractive contribution due to mass. 

 This question has relevance in the context of a black hole on the brane [3], 
which would be sitting in a trace free energy distribution coming as a 
backreaction of the bulk on the brane. The question is what should be the 
sign of this energy density? It has been argued that it must be negative if 
it is to strengthen black hole's gravity. \\

\n
 8.3 {\it Empty Space}
  
 In the Newtonian theory empty space is defined by vanishing of the matter 
density. Could we not define it similarly in GR at least for the simple case 
of space surrounding a static massive body? The analogue of the Newtonian 
density is the energy density measured by a static observer as defined by 
$\rho = T_{ab}u^au^b, u^au_a = 1$. In GR, we also have the null particles, and
 the energy density felt by them would be $\rho_n = T_{ab}k^ak^b, k^ak_a = 0$. 

 Shouldn't vanishing of these two densities be sufficient to characterize 
empty space for the situation under consideration? It turns out that it does 
[4]. The 
condition $\rho_n = 0$ for radially propagating null particles would imply 
$R^t_t = R^r_r$ which would lead to $AB = 1$. Putting this into the condition 
$\rho = 0$, equivalently $G^t_t = 0$ determines $A = 1 - 2M/r$, and so we 
obtain the Schwarzschild solution which characterizes empty space for 
spherical symmetry.

 Thus for spherically symmetric spacetime, empty space could be defined in 
line with the Newtonian theory by the absence of energy density felt by both 
timelike and null particles. This is a physically interesting 
characterization. \\

\newpage
 8.4 {\it Black Hole}

 Black hole is a body from which nothing can emanate out. There is a rigorous 
mathematical characterization involving sophisticated geometric constructs. 
Physically the most appealing definition is based on the consideration of 
escape velocity tending to the velocity of light. Over 150 years before the 
relativistic black hole was realised, Laplace had enquired when could light 
not escape a gravitating body? This was though simple argument and roughly 
pointing in the right direction. It is however an invalid (though valid in 
Laplace's time) application of a formula which happens to give the right 
result. The 
formula is invalid for $\frac{1}{2} m\, v^2$ is not the kinetic energy of a particle when 
its velocity tends to the velocity of light. Secondly, escape velocity being 
$c$ does not prohibit things to emanate from any surface while black hole is a
oneway surface, from which nothing can emerge.  

 I had been for long trying to devise a correct energetics argument for black 
hole. From what we have argued earlier, we could envision that gravitational 
charge, which could be defined in GR also for stationary field by 
appropriately adopting the Gauss theorem, is felt by the timelike 
particles through gravitational potential. On the other hand photons would 
feel the space curvature which is caused by the gravitational field energy. 
Photon is the limiting case of timelike particle. Now, what should happen 
when gravitational field energy  becomes comparable to gravitational charge. 
Where this happens timelike particles should be tending to null particles. 
This is an energetics consideration characterizing black hole by the property 
of ordinary particles tending to photons.

 The big question is how to get a measure of gravitational field energy which 
is notoriously ambiguous? There is a good prescription due to Brown and York 
[5] for total energy contained inside a sphere of radius $r$. By subtracting 
out mass at infinity from it we can get the measure of field energy which has 
the expected term, $-M^2/2r$ for large $r$. A black hole is 
characterized when gravitational charge coming from the Gauss theorem equals 
the field energy. 

 The Brown-York energy inside a sphere of radius $r$ is given by 
\be
E(r) = r - (g_{rr})^{-1/2}
\ee
and then the field energy would be given by
\be
E_F(r) = E(r) - E(\infty).
\ee

Our definition [6] for black hole is when 
\be
E_F = M_g = \frac{1}{4\pi}\int_{S^2}{\bf g.ds}
\ee
where $M_g$ is the gravitational charge and $\bf{g}$ is the red shifted proper
 acceleration as measured by an 
observer at infinity. It is the conserved mass $M$ for the Schwarzschild 
black hole while for the charged black hole it reads as $M - Q^2/r$ where $Q$ 
is the electric charge. It can be easily verified that our definition does 
give the right black hole 
radius for the spherical black holes. Further the characterizing relation has 
been rigorously established by using the Gauss-Codaci theorems and has 
also been applied to a black hole in the expanding universe [6].

 We have thus given a physically illuminating and intuitively appealing 
characterization of black hole based on an energetic consideration. It is 
quite insightful. Note that 
the condition $g_{tt} = 0$ is the statement equivalent to escape velocity 
being equal to velocity of light (it defines the static limit, a particle has 
to move with velocity of light to remain static) while the condition 
$g^{rr} = 0$ defines the confinement of photons. Since photons are 
always free and cannot be given arbitrary velocity, their confinement inside 
a surface means that its curvature should be so strong that photons can't 
propagate out and the surface becomes oneway. This is exactly what the latter 
condition indicates. \\

\n
 8.5 {\it Electrogravity Duality}

 It is the general property of a classical field to resolve it into its 
electric and magnetic parts. The manifestation of field due to a stationary 
source is called electric while that due to moving source magnetic. The terms 
electric and magnetic have been adopted because this resolution was for the 
first time realised for the electromagnetic field. One can transform from one 
to the other by a transformation which is given the name of duality 
transformation. 

 In the case of gravitation, the field is described by the Riemann curvature 
tensor which involves the second derivatives of the metric, and it thus has 
in fact dynamics of the field. In the case of electromagnetism, field 
contained the first derivative of the gauge potential, and dynamics was given 
by the field equation which is an independent statement involving second 
derivatives of the potential. Note that the field 
equation for gravitation is sort of implied by the Bianchi identity as shown 
above. We must therefore recognise the fact that gravitational field as 
described by the Riemann curvature is much subtler and higher order entity 
than electromagnetic field.

 In the context of electromagnetic resolution, the Riemann curvature is 
double 2-form, while the electromagnetic field is described by a single 
2-form. The projection of the 2-form and its dual onto a timelike unit vector 
$u^a$ of an observer gives respectively the electric and magnetic part. A 
dual is defined by $^*F_{ab} = \frac{1}{2}\eta_{abcd}F^{cd}$ where $F_{ab}$ is the 
electromagnetic field tensor and $\eta_{abcd}$ is the 4-dimensional 
volume element. The Riemann tensor $R_{abcd}$ has two 2-forms, the first two 
indices give the one and the last two the other. There would also occur two 
duals, left (on the first pair) dual and right (on the last pair) dual. To 
write the electric or magnetic parts each 2-form is to be projected on $u^a$; 
$E_{ab} = R_{acbd}u^cu^d$ is the electric part which is symmetric. The 
magnetic part would be given by $H_{ab} = {}^* R_{acbd}u^cu^d$ (projection of 
the right dual would give $H_{ba}$), which is trace free. Both of them would 
be orthogonal to $u^a$. $E_{ab}$ will account for 6 Riemann components of 
the kind $R_{\alpha t\beta t}$ and $H_{ab}$ will account for 8 components of 
the kind $R_{t\alpha\beta\gamma}$, and so there still remain 6 components 
unaccounted. These could be written as $\tilde E_{ab} = {}^ * R^*_{acbd} u^c u^d$, 
which would denote the purely spatial components, $R_{\alpha\beta\gamma\delta}$
. We would call $\tilde E_{ab}$ the passive electric part and $E_{ab}$ as the 
active electric part. 

 Gravitation has two kinds of sources, active the non-gravitational 
energy distribution and passive the gravitational field energy which give 
rise respectively to active ($E_{ab}$) electric part and passive 
($\tilde E_{ab}$) electric 
part. Note that we had argued earlier that the field energy produces spatial 
curvature which is denoted by the passive electric part. The magnetic part 
would be generated by motion of these sources. Note that electric parts are 
symmetric while magnetic part is not but is instead trace free. Thus any 
duality relation which does not impose any condition could only involve the 
electric parts. It turns out that if we consider the transformation involving 
all the three electromagnetic parts similar to the electrodynamics, the 
transformation would lead to the vacuum equation [7]. This has happened 
because the Riemann curvature does contain the dynamics of the field and the 
duality relations would prescribe certain relations between the curvature 
components giving the vacuum equation.

 In electrodynamics, the duality transformation is the symmetry of the vacuum 
equation, which in this case is given by
\be
E \, or \,  \tilde E = 0, ~H_{ab} = H_{ba}, ~E_{ab} + \tilde E_{ab} = 0.
\ee
This would be invariant for the transformation,
\be
E_{ab}\leftrightarrow \tilde E_{ab}, ~H_{ab}\rightarrow H_{ab}
\ee                                                             
which we call the electrogravity duality transformation [7]. The scalar 
curvature 
is given by $R = 2(E - \tilde E)$, which would change sign under duality. And 
so would the Weyl curvature. The duality transformation would imply 
$GM\rightarrow - GM$ in the Schwarzschild solution. This happens because the 
duality means interchange of active and passive parts which are respectively 
anchored to positive non-gravitational matter/energy and to negative 
gravitational field energy.

 In the familiar terms duality is equivalent to the interchange of the Ricci 
and Einstein tensors. This is because contraction of Riemann is the Ricci 
while that of its double dual is the Einstein tensor. The duality is 
the interchange between Riemann and its double dual. Thus at the level of 
equations, we just need to interchange Ricci and Einstein. The vacuum 
equation is however insensitive to this interchange and remains invariant. 
This is not so for non empty space. The duality transformation would for a 
perfect fluid imply $\rho\rightarrow \frac{1}{2}(\rho + 3p), p\rightarrow 
\frac{1}{2}(\rho - p)$ . The de Sitter space is dual to anti de Sitter, while the 
radiation universe would be self-dual.

 Though the vacuum equation is invariant under duality, the effective vacuum 
equation as considered in 8.3 above, $\rho = 0, \rho_n = 0$ is not. For 
spherical 
symmetry, the Schwarzschild solution is unique and hence all the equations 
admitting it as general solution would equally well characterize vacuum. Under 
the duality transformation, we have  $\rho\rightarrow\rho_t, \rho_n
\rightarrow\rho_n$ where $\rho_t = (T_{ab} - \frac{1}{2}T g_{ab})u^au^b$ is the 
timelike convergence density.

 The dual equation to the effective vacuum equation would be $\rho_t = 0, 
\rho_n = 0$ which in terms of the familiar Ricci components would read as 
$R^t_t = R^r_r = 0$. It would solve to give $A = 1/B = 1 + 2\phi, \phi = 
-k - M/r$. That is it restores the additive constant in the potential as 
was the case for the Newtonian theory. The spacetime would be non empty and 
would have the stresses, $T^t_t = T^r_r = 2k/r^2$. In the Newtonian limit, it 
would not be distinguishable from the Schwarzschild solution. The dual 
solution thus has the basic features of the Schwarzschild solution but it is 
not asymptotically flat. To seat the Schwarzschild particle in a realistic 
situation, we have to break asymptotic flatness so as to allow for existence 
of other matter in the Universe without affecting the basic character of the 
field. This is precisely what has been achieved in the dual solution. The 
dual solution is thus Machian in spirit and is closer to the realistic 
setting.

 This is also the Barriola - Vilenkin solution [8] describing asymptotically 
the field of a global monopole. Global monopole is a topological defect which 
is supposed to arise when global $O(3)$ symmetry is spontaneously broken into 
$U(1)$ in the phase transitions in the early Universe. Thus the duality 
transformation implies putting on a global monopole on the Schwarzschild 
particle.   

 Similarly the solutions dual to the NUT space, charged black hole and Kerr 
black hole have been considered. Further the applications of the duality 
to the dilaton gravity as well as in in higher and lower dimensions have been 
studied. In particular, an interesting new black hole spacetime has been 
obtained as a solution of the dual equation in 2+1 gravity [9]. It has also 
been shown that the solution of the static massless minimally coupled scalar 
field is also dual to the Schwarzschild field. This is obtained by writing 
$\rho_n$ for, instead of radial, transversely propagating photons [10]. 
There could therefore occur more than one dual solutions, however they would 
all agree with the original solution in the first approximation. \\

\n
{\bf  9. End of everything}

 The breakdown of theory is called by the beautiful name of singularity. It 
indicates a situation where physical parameters of the theory become untenable
 by attaining infinite values. For GR, it is the spacetime curvature as well 
as density and pressure that should blow up at the singularity. The spacetime 
curvature becoming infinite indicates the breakdown of spacetime structure. 
The singularity in GR not only marks the end of GR, but also of everything 
else. No other physical theory or structure can survive in absence of the 
proper spacetime background. Thus like gravity its singularity is also 
universal.

 There are powerful singularity theorems due to Penrose and Hawking that prove 
inevitability of occurrence of singularity in GR under quite general 
assumptions except one, the occurrence of compact trapped surfaces. A trapped 
surface essentially defines photon confinement inside a compact surface. The 
occurrence of trapped 
surface should rather be governed by the field equation. Taking this as an 
assumption is of course at non-trivial cost of generality which is the 
hallmark of the theorems. The influence of the
theorems was however so strong and wide spread that the occurrence of 
singularity in GR is 
inevitable became a folklore. The observation of cosmic microwave background 
radiation implying the big-bang singular beginning of the Universe 
strengthened it immensely. This view was not correct was demonstrated by 
Senovilla in 1990 when he obtained a singularity free cosmological model 
filled with radiation with all physical and kinematic parameters remaining 
finite and regular everywhere [11]. The theorems became inapplicable because 
it did not satisfy the assumption of trapped surface. It was also a 
demonstration of the fact that violation of this assumption entails no 
unphysical features. The Senovilla class of models were cylindrically 
symmetric. Since then spherically symmetric imperfect fluid models having no 
singularity have been obtained, including the one which is also 
oscillating [12]. It oscillates between two regular states without 
ever becoming singular anywhere. That is there do exist truly non-singular 
cosmological solutions in GR. It is another matter whether they could 
be applied to the actual Universe.

 One of the most important unresolved questions is, what is the ultimate end 
product of gravitational collapse? There is no question that it would end 
into a singularity as trapped surface would always form in this case. The 
question is, will it always be hidden behind a horizon 
(black hole) or sometimes visible (naked) to asymptotic observer? Penrose has 
proposed the cosmic censorship hypothesis to say that there exist no naked 
singularity in the Universe.  
The theorems only guarantee existence of singularity but give no clue about 
its nature and visibility to external world. There exists no proof or 
otherwise of the hypothesis. A considerable effort has been invested in 
studying the question in various settings (see a recent review [13]). It is 
perhaps now generally agreed that there 
exists regular initial data set which could lead to black hole as well as to  
naked singularity. It is being argued that cause for occurrence of naked 
singularity is 
inhomogeneity of density and presence of shear which make collapse incoherent 
and thereby delay formation of trapped surfaces. The race is between 
formation of 
singularity and trapped surface. If the latter forms first then it is black 
hole while if the former forms first then it would be visible thus naked to 
external observer before it is ultimately engulfed by the horizon. It appears 
that shear plays very important role in this process.

 One of the ingenious applications of naked singularity is the proposal for 
the source of gamma ray bursts [14]. In practical terms we could model naked 
singularity as a seat of divergingly high curvatures. This could occur 
quite naturally in a massive collapsing object. The ultra high curvatures 
could produce a fireball giving rise to shocks in the on falling matter and 
ultimately result in producing high energy gamma rays. It happens for a very 
brief period before the trapped surface sets in and the object becomes a black 
hole. In 
this proposal gamma ray bursts could be envisioned as birth-cries of black 
holes. The crucial question is formation of singularity before setting in of 
the trapped surface which is delayed by presence of inhomogeneity and shear. 
Remarkably, this proposal is based purely on classical physics and invokes no 
new and speculative phenomena.  \\

\n 
{\bf  10. Looking ahead}

 Gravity is a universal interaction and it is therefore not a force like 
other forces but is a property of spacetime structure of the Universe. The 
question is how do we address the singularity in the structure of spacetime, 
indicating its breakdown not only for description of gravity but also as 
background for all physical phenomena. It indicates the end of everything, 
and should the answer of it be the Theory of Everything? So do the proponents 
of the string theory claim. The string theory is one of the proposals for a 
covering theory to 
GR which is vigorously being pursued. It proposes in principle 
a grand vision of synthesis of all forces, quantization of gravity and origin 
of matter in the Universe. All this is supposed to happen in 10 dimensions. 
The extra space dimensions are supposed to be compact and manifest only at a 
very high energies. GR would be the low energy limit of the theory. There is 
however no natural and unique way to come from 10 to 4 dimensions. 

 The other approach is that of canonical quantum gravity a la Abhay Ashtekar 
et al. How to tailor spacetime geometry so that it becomes 
amenable to quantization? It is essentially the quantization of spacetime 
geometry. New mathematical tools and constructs as well as new concepts are 
being developed so as to handle discretization of inherently continuum 
structure of spacetime. They are in fact evolving quantum geometry. This 
proposal remains confined to the familiar 4-dimensions.  

 Both the approaches have about equal share in success, for instance in 
understanding black hole entropy to some extent. Yet they have both long 
way to go. The string theory has a strong backup of the large particle physics 
community while the canonical qauntization is pursued mainly by a small 
group of relativists. The two approaches are quite different, the former 
relies on the field theoretic techniques and concepts while the latter on the 
geometric constructs. The two will have to converge as they asymptotically 
approach the ultimate theory of spacetime, which may or may not be a theory 
of everything! At any rate, one thing is certain that we do need a new theory 
which could be string theory or canonical quantum gravity, or something else! 

 Let us try to see into the unknown on the basis of wisdom gained from our 
past experience. From Newton to Einstein, the guiding force was light, its 
constant speed and its interaction with gravity. The latter required space 
to be curved. On the basis of these simple facts, we could argue that 
gravitational potential should not only give the acceleration to ordinary 
particles but also curve space for photons to feel gravity. Since acceleration 
is given by gradient, constant potential has no dynamical effect. But it is 
not so for space curvature, constant potential does produce non-zero 
curvature as shown above in 8.1. The journey from old to new theory is always 
of synthesis. For description of the field of a mass point, one cannot tamper 
with the Laplace equation of the Newtonian theory because of its strong 
observational support. At the same time, the potential must also curve the 
space which would require its absolute determination. That it can vanish only 
at infinity and nowhere else. Thus the GR in contrast to the Newtonian theory 
determines the potential absolutely. This is a synthesis.

 Further, as there exists the escape velocity threshold for timelike 
particles, there should also exist a similar threshold for photons. Photons 
cannot turn back like ordinary particles, the only way they could be kept 
bound to a gravitating body is that they are not let to propagate out of a 
compact surface. That would happen when the surface becomes null;i.e. its 
normal is null. It then becomes a oneway membrane defining the horizon of a 
black hole. The existence of such a surface therefore becomes a natural 
requirement the moment we have photons to contain. These important and 
distinguishing features of GR could be deduced in principle without reference 
to the full theory. 

 The point I wish to make is that it should always be possible to anticipate 
and deduce some of the features of the new theory. Now when we wish to go 
beyond Einstein in the high energy regime, what kind of new features could one 
expect? Unfortunately, there does not seem to be anything like light showing 
us the way. One possible entity for synthesis is gravitational field energy 
like the gravitational potential in the Newton-Einstein synergy. In GR, it has 
no associated stress tensor and only comes through the space curvature. In the 
high energy limit, could it happen that it becomes concrete like other matter 
fields and attains a stress tensor? Something like this does happen in the 
string theory inspired brane world model [15]. In this model, all matter 
fields are supposed to remain confined to the 4-dimensional spacetime, called 
the brane while gravitational field can propagate in the 5-dimensional 
spacetime, called the bulk. The bulk is however taken as  
the vacuum with negative cosmological constant. Also is assumed the 
reflection symmetry in the extra space dimension. Then it turns out that it is 
not necessary to restrict the extra dimension to be finite and compact. By 
this construction, one of the remarkable things that happens is that the free 
gravitational field from the bulk gets ``reflected'' as backreaction onto  
the brane as trace free matter field. This is something in line with what one 
can extrapolate for a new synthesis. There is currently a good 
degree of activity in the brane world model.

 The measure of a theory is the kind of questions it admits. New theory admits
questions that were not admissible in the earlier theory. For instance, it is 
a valid question to ask in GR, when did the spacetime (the Universe) begin? 
And the answer is that it had its explosive birth in hot big-bang about 
15-20 billion years ago. The next question in this series for the new theory 
to answer would be, what is the spacetime made of? That is the fundamental 
question which should be answered by the new theory.
 
 A fundamental theory also has an impact on the world view. The view that has 
gained acceptance amongst people at large. For instance, the fact that the 
earth is not flat but is a curved surface like a ball has been internalized 
and assimilated in the knowledge base of the present day society. At a slightly
 higher degree of abstraction, the fact that material bodies attract 
each-other by an invisible force of gravity has also descended down to the 
common knowledge culture. The next step of advancement in the similar vein 
would be the assimilation of the fact that massive bodies curve the space 
around them. That is the abstract spacetime manifold (the Universe) we live 
in is not flat but is curved. One of the ways to measure this curvature is by  
measuring  bending of light when it grazes a massive body like a star. Thus 
the geometry of spacetime should be of as general a concern as the geometry 
of the planet earth we live on. The only difference is that the former is an 
abstract entity while the latter is the concrete rock and sand. 

 Like gravity social interaction is also universal and of great consequence 
for our existence and well being. It should thus also attract collective 
concern and attention of all of us with utmost seriousness and commitment. We 
should never forget the fellow citizens who have been contributing from their
honest earnings for our upkeep as well as for the facilities we use for our 
work. 
Apart from contributing to the knowledge base by our work in our specific 
discipline, as persons of learning and more importantly practitioners of 
scientific method we also owe to the society a studied and responsible 
participation in the discourses on the issues of wider social relevance. 

 Let us try to emulate the basic gravitational property of interaction with 
one and all, and come closer. If this is realised in the true spirit of both 
science and society, there is a good reason for hope for harmony, peace and 
an enlightened world community. This is a matter of gravity for one and all, 
once again a universal pronouncement of profound significance and value.

 I close by quoting Ghalib, the great Urdu poet of the 19th century.

\vspace{1cm}

\begin{center}
 {\it There are many poets in the world great and competent} 
\end{center}
\begin{center}     
{\it yet they say different is Ghalib's manner.} 
\end{center}

\vspace{1cm}

\n
{\bf Acknowledgement:} Learning is a long and cooperative process in which 
contributions come from various quarters consciously as well as unconsciously.
Apart from my doggedness to understand things in the farmer's robust common 
sense way, the exercise of giving popular lectures to audiences of varying 
background has played very important role in devising new and novel 
arguments. I have learnt a lot from  this activity as well as from large 
number of my collaborators, too many to name, from far and wide. Of late it 
has been my student, Parampreet who has borne admirably the bombardment of 
crazy and stupid ideas, and has thus served as a good sounding board, and 
hence I can't help naming him.
 \newpage

\newpage


\begin{thebibliography}{}
\bibitem{}
 P Singh \& N Dadhich, gr-qc/9912028, to appear in {\it Int. J. Mod. Phys. A} 
\bibitem{}
 N Dadhich (2000) {\it Phys. Lett.} {\bf B492}, 537
\bibitem{}
 N Dadhich, R Maartens, P Papadopoulos \& V Rezania (2000) {\it Phys. Lett.}\\
\hspace*{0.1cm} {\bf B487}, 1
\bibitem{}
 N Dadhich (2000) {\it Current Science} {\bf 78}, 1118
\bibitem{}
 J D Brown \& J W York (1993) {\it Phys. Rev.} {\bf D47}, 1407
\bibitem{}
 N Dadhich (1999) {\it Current Science} {\bf 76}, 831;\\
\hspace*{0.1cm} S Bose \& N Dadhich (1999) {\it Phys. Rev.} {\bf D60}, 064010(7)
\bibitem{}
 N Dadhich (2000) {\it Gen. Relativ. Grav.} {\bf 32}, 1009\
\bibitem{}
 M Barriola \& A Vilenkin (1989) {\it Phys. Rev. Lett.} {\bf 62}, 341
\bibitem{}
 S Bose, N Dadhich \& S Kar (2000) {\it Phys. Lett.} {\bf B477}, 451
\bibitem{}
 N Dadhich \& N Banerjee, hep-th/0012015
\bibitem{}
 J M M Senovilla (1990) {\it Phys. Rev. Lett.} {\bf 64}, 2219;\\
\hspace*{0.3cm}  E Ruiz \& J M M  
Senovilla (1992) {\it Phys. Rev.} {\bf D45}, 1995
\bibitem{}
 N Dadhich (1997) {\it J. Astrophys. Astr} {\bf 18}, 343;\\
\hspace*{0.3cm} N Dadhich \& A K Raychaudhuri (1999)
{\it Mod. Phys. Lett.} {\bf A14}, 2135
\bibitem{}
 P S Joshi (2000) {\it J. Astrophys. Astr} {\bf 55}, 529
\bibitem{}
 P S Joshi, N Dadhich \& R Maartens (2000) {\it Mod. Phys. Lett.} {\bf A15}, 991
\bibitem{}
 L Randall \& R Sundrum (1999) {\it Phys. Rev. Lett.} {\bf 83}, 4690; \\
 \hspace*{0.3cm}  R Maartens (2000) {\it Phys. Rev.} {\bf D62}, 084023
\end{thebibliography}
\end{document}